\documentclass[letterpaper,11pt]{article}
\usepackage{amsmath,amssymb}
\usepackage{bm}
\usepackage{graphicx,epsfig,psfrag}
\usepackage{cite}

\makeatletter
 
 \@addtoreset{equation}{section}
\makeatother

\newcommand{\gsim}{\mathrel{\lower2.5pt\vbox{\lineskip=0pt\baselineskip=0pt
                   \hbox{$>$}\hbox{$\sim$}}}}
\newcommand{\lsim}{\mathrel{\lower2.5pt\vbox{\lineskip=0pt\baselineskip=0pt
                   \hbox{$<$}\hbox{$\sim$}}}}

\newcommand{\sla}[1]{{\raise.15ex\hbox{$/$}\kern-.57em #1}}
\newcommand{\Sla}[1]{\kern0.12em{\raise.15ex\hbox{$/$}\kern-.74em #1}}

\newcommand{\beq}{\begin{eqnarray}}
\newcommand{\eeq}{\end{eqnarray}}
\newcommand{\nn}{\nonumber}

\begin{document}


\begin{titlepage}

\setcounter{page}{0}

\begin{flushright}
EDINBURGH-2008/05
\end{flushright}

\vskip 3cm

\begin{center}

{\huge\bf A composite gluino at the LHC}

\vskip 1cm

{\large {\bf Thomas Gr\'egoire}$^a$ and {\bf Emanuel Katz}$^b$}

\vskip 0.5cm
$^a${\it SUPA, School of Physics, University of Edinburgh, \\
               Edinburgh, EH9 3JZ, Scotland, UK} \\
$^b${\it Physics Department, Boston University,\\
         590 Commonwealth Avenue, Boston, MA 02215, USA}

\abstract{
We investigate the decay of particles with the quantum numbers 
of the gluino.  Besides SUSY, such particles may be present
in models where the Higgs and top are composite.  We find
that such 'composite' gluinos have decay signatures similar 
to those of gluinos in 'more minimal' SUSY type models.
Though it is in principle possible to distinguish the two
scenarios, we find that it will be a challenging task
at the LHC.  This puts into question the common lore
that a gluino is an obvious 'smoking-gun' signature of SUSY.
}

\end{center}
\end{titlepage}


\section{Introduction}
\label{sec:intro}

The Large Hadron Collider (LHC) promises to give us a glimpse of the physics
behind electroweak symmetry breaking.  If our ideas about naturalness are
correct, we should expect new colored particles whose decays provide
exciting signals.  Once these new particles are discovered, the next task
will be to describe the particulars of the signal within some framework
for physics beyond the Standard Model.  An interesting question, that
broadly distinguishes between various scenarios is whether there is any
hint of compositeness in the signals.  Bounds on higher dimension operators
suggest the compositeness scale for the first two generations should be
too high to be relevant at the LHC.  On the other hand, less is known 
about the third generation.  In models where the 
Higgs is composite, such as the Little Higgs \cite{ArkaniHamed:2002qy,ArkaniHamed:2001nc,Katz:2003sn}, having a partly composite top
and bottom is well motivated.  Indeed, the little hierarchy, as
well as the large top Yukawa coupling to the composite Higgs, suggest that the top, 
and perhaps the left-handed bottom, are composite.   Thus, there may be 
different TeV suppressed operators that provide new interactions for the
bottom and the top.  These might be difficult to measure in a direct way
experimentally.  There might, however, be a different way to look
for compositeness.  Namely, if there are new composite states which are
accessible at the LHC, then these will generally have non-renormalizable
interactions with tops and bottoms.  If these interactions 
dominate the decay of the new particles, then by studying this decay
one could infer the presence of compositeness.  At the LHC this
seems especially promising if some of the new particles are colored.

One would like to distinguish composite Higgs models from
supersymmetry (SUSY).   After all, SUSY is the main viable framework 
where the Higgs and other Standard Model particles are fundamental.
The trouble is that composite Higgs models with a composite top 
could easily have TeV mass fermions which carry the quantum
numbers of SUSY gauginos and Higgsinos\cite{Katz:2003sn}.  In addition, there
can be a discrete symmetry like R-parity which insures that
the lightest of these is stable, and thus that any decay of the 
others is accompanied by missing energy.  The strong dynamics
that is behind the composite Higgs models implies that these new 
composite fermions will couple to tops and bottoms through 4-fermi 
interactions.  In Little Higgs theories, these interactions are
expected to be suppressed by a scale of order 1 TeV.
One thus obtains decay chains for composite gauginos similar to
those in 'more minimal' SUSY models (MMSSM), where 
all scalar partners, except for the stops 
and left handed sbottom, are heavy.  

How then should we tell apart SUSY from the composite models?
A particularly attractive particle to study, in that regard, is the gluino (see \cite{Lillie:2007hd,Georgi:1994ha} for generic effects of a  composite top only).  
Since it is an adjoint of color and is also a fermion, it
should be one of the more abundantly produced particles at
the LHC, and has a good chance of being discovered. 
In both the MMSSM and composite cases it will
decay into tops, bottoms, and W's with missing energy 
carried away by the stable particle (which we will assume
is a neutralino).  To differentiate SUSY from the the composite
case one therefore needs to look beyond just the final states.
In this paper we will focus on the case where the dominant
decay of the gluino is into a top, a bottom, and a chargino 
(which subsequently decays into a neutralino via W emission).
Thus, the only invariant to consider, sensitive to the 
nature of the gluino, is the top-bottom invariant mass, $m_{tb}^2$.
Fortunately, due to the global symmetry of the composite
model, the $m_{tb}^2$ invariant distribution has features which
would be absent in most MMSSM models.  We investigate 
whether it would be possible to use this distribution 
to distinguish SUSY from non-SUSY gluinos at the LHC.  We
find that, generically, due to the combinatorical background,
identifying the physics behind the distribution
may be difficult.  It is therefore far from clear that
the gluino is an obvious signal of an underlying SUSY theory
(see \cite{Alves:2006df,Wang:2006hk,Csaki:2007xm} for study on how to distinguish a gluino from a vector particle with the same quantum numbers).

As it is challenging to tell the composite model apart from 
the SUSY case based on the gluino alone, we also briefly consider the
possibility of finding evidence for stops (in the MMSSM context).
Namely, we consider the case that the stops are light enough
to be directly produced, and then decay to a bottom and a chargino.
Here, we look for an excess of events, by choosing cuts which 
reduce both Standard Model top production, as well as
products from the gluino decay.  We find that for a light stop,
and a heavy gluino, isolating the stop should be possible.

The paper is organized as follows.  In section~\ref{compferm}
we review why in models with a composite top/Higgs we 
expect other composite 'gaugino' fermions and describe 
their interactions with tops and bottoms.  We then 
contrast these interactions with those present in SUSY,
and explain how they affect the $m_{tb}^2$ observable
in section~\ref{mtb}.  In section ~\ref{results}
we present our parton-level simulation of the distributions
in the SUSY and composite cases for various parameter values.  
We include some detector effects, and provide the resulting plots, 
after appropriate cuts.  We consider the case of 
having a light stop in section ~\ref{stop}, and
conclude in section ~\ref{end}.

\section{Composite Gauginos}
\label{compferm}

Let us first recall why composite fermions with the quantum numbers
of gauginos may be generic in models where the top and Higgs are
composite(see \cite{Katz:2003sn} for a concrete model).  If we assume that the underlying strong dynamics
behind the compositeness is non supersymmetric, then it is
reasonable that both the Higgs and the top are bound states
of some new fundamental fermions.  The simplest guess is that
the Higgs, $h$, is a bound state of two of these fermions:
\beq
h \sim (\Psi_{2_W} \Psi_0),
\eeq
an $SU(2)_W$ doublet fermion, $\Psi_{2_W}$, and an $SU(2)_W$
singlet fermion $\Psi_0$.  To address the little hierarchy problem,
and to motivate why the top has such a large coupling to the Higgs,
it is natural that at least the right handed top, $t^c$, and possibly
the left handed top and bottom, $q_L$, mix with heavy composite vector fermions 
$T^c$ and $Q_L$, of similar quantum numbers.  These composite states are likely
to be bound states of three fundamental fermions:
\beq
&& T^c \sim (\Psi_{\bar{3}_c} \Psi_0 ~\Psi_0') \nn \\
&& Q_L \sim (\Psi_{3_c} \Psi_{2_W} \Psi_0'),
\eeq
where $\Psi_{3_c}$ and $\Psi_{\bar{3}_c}$ are in the fundamental and anti-fundamental
representation of color, and $\Psi_0'$ is a Standard Model singlet.
In Little Higgs models, for example, $T^c$ and $Q_L$, are the top partners
which protect the Higgs from getting a quadratically divergent contribution to
its mass from top loops.  The point is that if the underlying strong dynamics
allows the above fermion bound states, then it also allows for 
'gluinos', 'winos', 'binos', and 'higgsinos'
\beq
&& \tilde{g} \sim (\Psi_{3_c} \Psi_{\bar{3}_c} \Psi_0') \nn \\
&& \tilde{\omega}, \tilde{B}  \sim (\Psi_{2_W} \Psi_{2_W} \Psi_0') \nn \\
&& \tilde{H} \sim  (\Psi_{2_W} \Psi_0 ~\Psi_0'). 
\eeq
Moreover, since from the point of view of the strong dynamics, all the $\Psi$'s
are the same, the masses of these 'gauginos' should be of the same order 
as the masses of the top partners.  The latter cannot be much heavier
than a TeV, for that would imply a fine tuning in the mass of the Higgs.
Consequently, the masses of the composite 'gauginos' should also be around
a TeV. Finally, we note that, as in the MSSM, we can introduce in this sort of model an R-parity:$(-1)^R = (-1)^{3B+L +2S}$ under which the 'gauginos' are odd and the Standard Model particles  even. This results in a stable 'gaugino'  \cite{Katz:2003sn}.

We will focus on the 'gluino', and so let us now describe its interactions with 
tops and bottoms.  As is the case for nucleons at energies below $\Lambda_{QCD}$,
our composite fermions will have all manner of 4-Fermi interactions consistent
with symmetries.  For the 'gluino' the relevant ones are
\beq
L &= &c_1~\frac{(\tilde{g} ~q_L)(q_L ~\tilde{\omega})^\dagger}{f^2} 
+ c_2~\frac{(\tilde{g} ~q_L)(q_L ~\tilde{B})^\dagger}{f^2} \nn \\
&+& c_3'\left(\frac{(\tilde{g} ~q_L)(t^c ~\tilde{H})}{f^2} + \frac{(\tilde{g} ~t^c)(q_L ~\tilde{H})}{f^2}\right)
+ c_3~\frac{(\tilde{g} ~\tilde{H})(q_L ~t^c)}{f^2}  
+ c_4~\frac{(\tilde{g}~\tilde{H})^{\dagger} (q_L~t^c)}{f^2} +  h.c.,
\eeq
where $q_L$ is the third generation quark doublet, $t^c$ is the right handed top, and
we use the notation $(\psi~\chi) = \epsilon^{\alpha\beta} \psi_{\alpha}\psi_{\beta}$
for the scalar contraction of two-component spinors.  
Note, that the two interactions proportional to $c_3'$ have identical strengths.  This follows again from the fact
that the strong dynamics allows for a global symmetry which rotates the $\Psi$'s amongst
each other and hence treats all composite fermions equally.  In fact, the identity
$\epsilon_{\alpha\beta}\epsilon_{\gamma\delta} + \epsilon_{\alpha\gamma}\epsilon_{\delta\beta}=
\epsilon_{\alpha\delta}\epsilon_{\gamma\beta}$ means that the $c_3'$ and $c_3$ terms are
equivalent.  We can therefore choose $c_3'=0$ without any loss of generality.
Once electroweak symmetry is broken, the 'chargino' and 'neutralino' spectrum and mixing
angles, together with the $c_i$ will determine the dominant decay mode of the 'gluino'.  
We will assume that a 'gluino' will mostly decay into a chargino with a significant
higgsino component.  Therefore, we will only need the interactions
\beq
L^{\text{comp}}_{gbtC} = c_3 ~\frac{(\tilde{g} ~C^+)(b ~t^c)}{f^2} +c_4~\frac{(\tilde{g} ~C^-)^\dagger(b ~t^c)}{f^2} 
+ h.c.
\eeq
Before continuing, let us observe that imposing the charge conjugation symmetry, $t^c \rightarrow b$, on the possible
4-Fermi interactions (which for us followed from the larger global symmetry), leads to operators where the top and
bottom spinors must be contracted with each other. This will be important later on, as it implies that resulting
matrix elements must be proportional to $2 p_t \cdot p_b \sim m_{tb}^2$ (the top-bottom invariant mass).
  
In a supersymmetric model, the gluino could decay through similar operators if the stop and sbottom are heavier 
than the gluino:
\beq
L^{\text{susy}}_{gbtC} = c_t~\frac{(\tilde{g} ~t^c)(b ~C^+)}{f^2} + c_b~\frac{(\tilde{g} ~b)(t^c ~C^+)}{f^2} + h.c.
\eeq
where the first operator corresponds to the exchange of a heavy right-handed stop, while the second corresponds 
to the exchange of a left-handed sbottom. Notice that there are no operators of the form $c_3, c_4$ as those 
would require the exchange of a charged color octet, which is absent in the MSSM.  The above charge conjugation 
symmetry would impose $c_t = c_b$, and would again force the top and bottom momenta to be contracted.  However,
for a generic SUSY spectrum, we do not expect this symmetry and therefore we will not consider this case further.

\section{The $m_{tb}^2$ invariant}
\label{mtb}
\begin{figure}[h]
\begin{center}
\psfrag{Labelx}{$m_{tb}^2$($10^5 \text{GeV}^2$)}
\psfrag{Labely}{\begin{tabular}{c}$m_{tC}^2$\\ ($10^5 \text{GeV}^2$)\end{tabular}}
\psfrag{Labelz}{\begin{tabular}{l}$\mathbf m_{ \tilde{\mathbf g}} \mathbf{ = 1}$ \bf TeV \\ $\mathbf m_{\mathbf C^+} \mathbf {= 300}$ \bf GeV\end{tabular}}
\includegraphics[width=15cm]{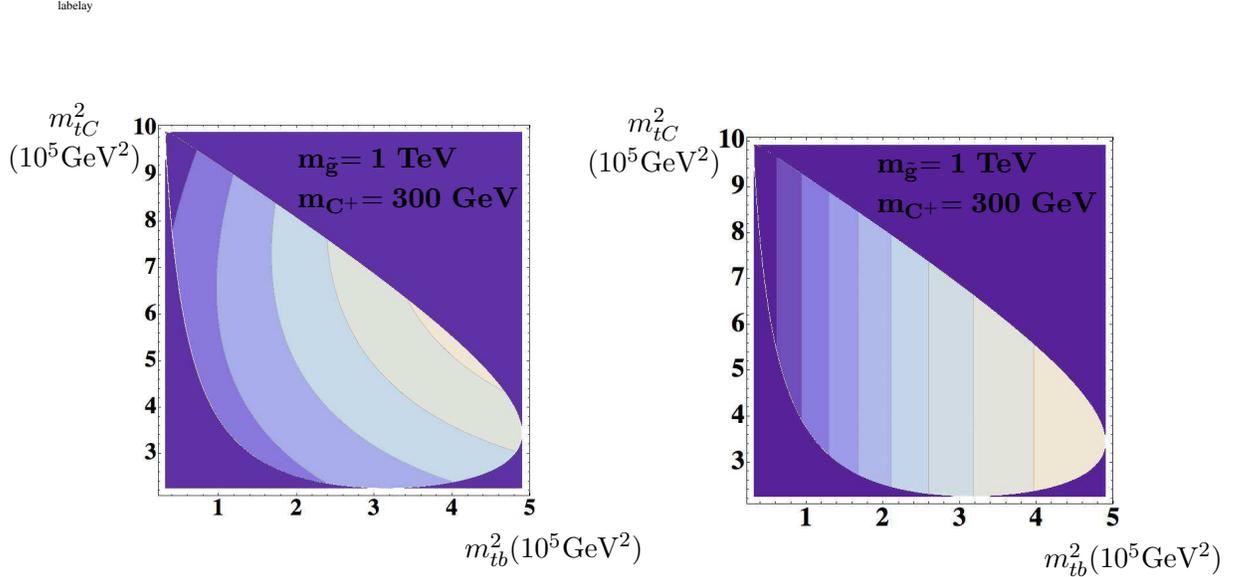}
\caption{Dalitz plot for $c_t=1,c_b=2,c_3=c_4=0$ (left) and $c_3=1,c_t=c_b=c_4=0$(right).}
\label{fig: dalitz}
\end{center}
\end{figure}

To distinguish between the composite and supersymmetric models, we study the structure of the gluino decay. 
The different operators lead to  different distributions for the invariant masses $m_{tb}^2 = (p_t + p_b)^2$ 
and $m_{t C}^2 = (p_t + p_{C})^2$. In figure \ref{fig: dalitz} we show the Dalitz plot for the gluino decay 
to $t,b,C$ for different choices of $c_t,c_b,c_3,c_4$. 
We notice that in the case of the composite gluino, the differential width $d\Gamma/(dm_{tb}^2 dm_{t\tilde{C}}^2)$ depends only on the invariant $m_{tb}^2$. 
This follows from momentum conservation, and our earlier observation that top and bottom momenta must be contracted in the matrix element.
Unfortunately, we cannot use  this feature to distinguish the two scenarios because the chargino subsequently decays to missing energy, 
and the invariant mass $m_{t C}$ cannot be measured.

\begin{figure}[h]
\begin{center}
\psfrag{SPECTRUM}{\begin{tabular}{l} $m_{\tilde{g}}=1$ TeV \\ $m_C = 300$ GeV \end{tabular}}
\psfrag{SPECTRUM1}{\begin{tabular}{l} $m_{\tilde{g}}=1$ TeV \\ $m_C = 100$ GeV \end{tabular}}
\psfrag{SPECTRUM2}{\begin{tabular}{l} $m_{\tilde{g}}=1$ TeV \\ $m_C = 600$ GeV \end{tabular}}
\psfrag{LABELX}{$m_{tb}^2$ ( $10^5 \text{GeV}^2$ )}
\includegraphics[width=15cm]{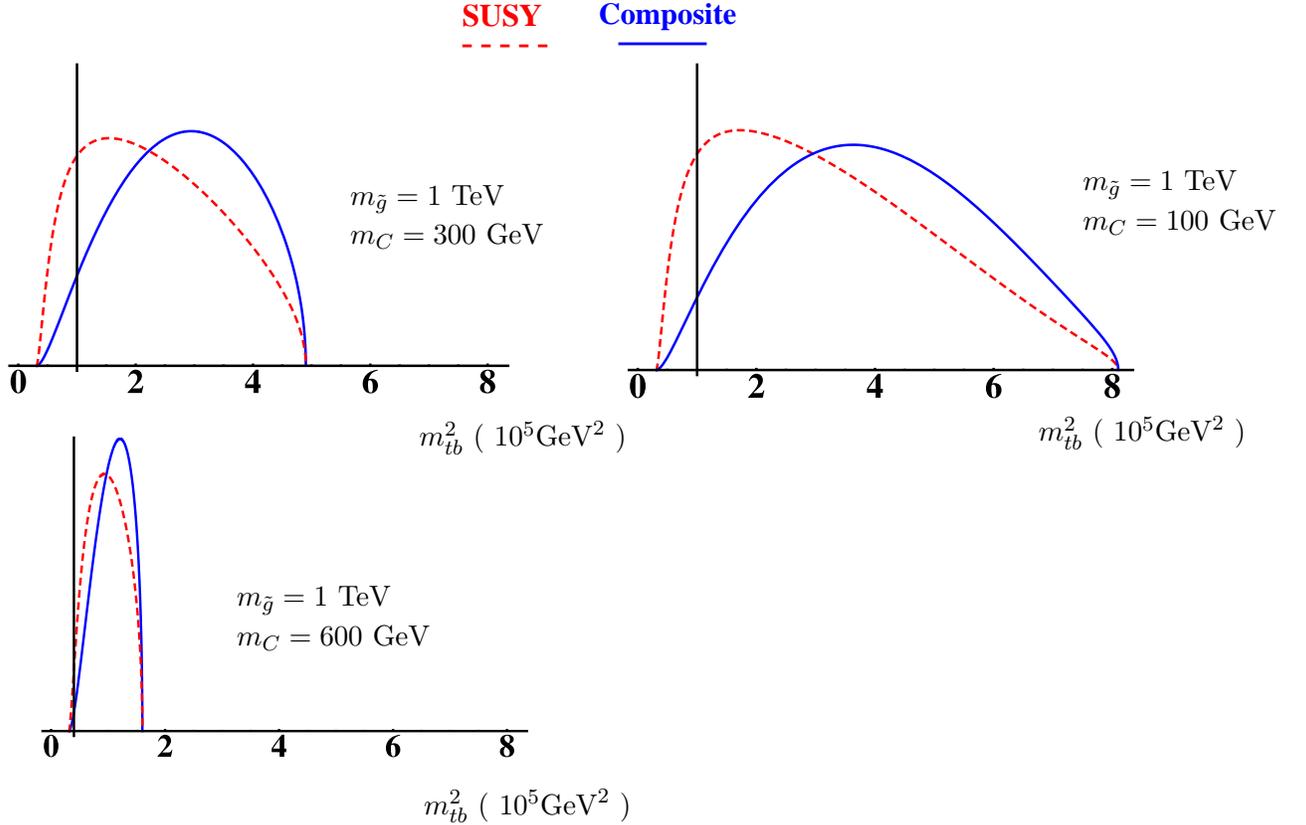}
\caption{$m_{tb}^2$ distribution for the operator $c_t$(in blue) and $c_{3}$(in red) for various spectrum}
\label{fig: mtbtheory}
\end{center}
\end{figure}

\begin{figure}[h]
\begin{center}
\psfrag{SPECTRUM}{\begin{tabular}{l} $m_{\tilde{g}}=1$ TeV \\ $m_C = 300$ GeV \end{tabular}}
\psfrag{SPECTRUM1}{\begin{tabular}{l} $m_{\tilde{g}}=700$ GeV \\ $m_C = 10$ GeV \end{tabular}}
\psfrag{SPECTRUM2}{\begin{tabular}{l} $m_{\tilde{g}}=800$ TeV \\ $m_C = 100$ GeV \end{tabular}}
\psfrag{LABELX}{$m_{tb}^2$ ( $10^5 \text{GeV}^2$ )}
\includegraphics[width=15cm]{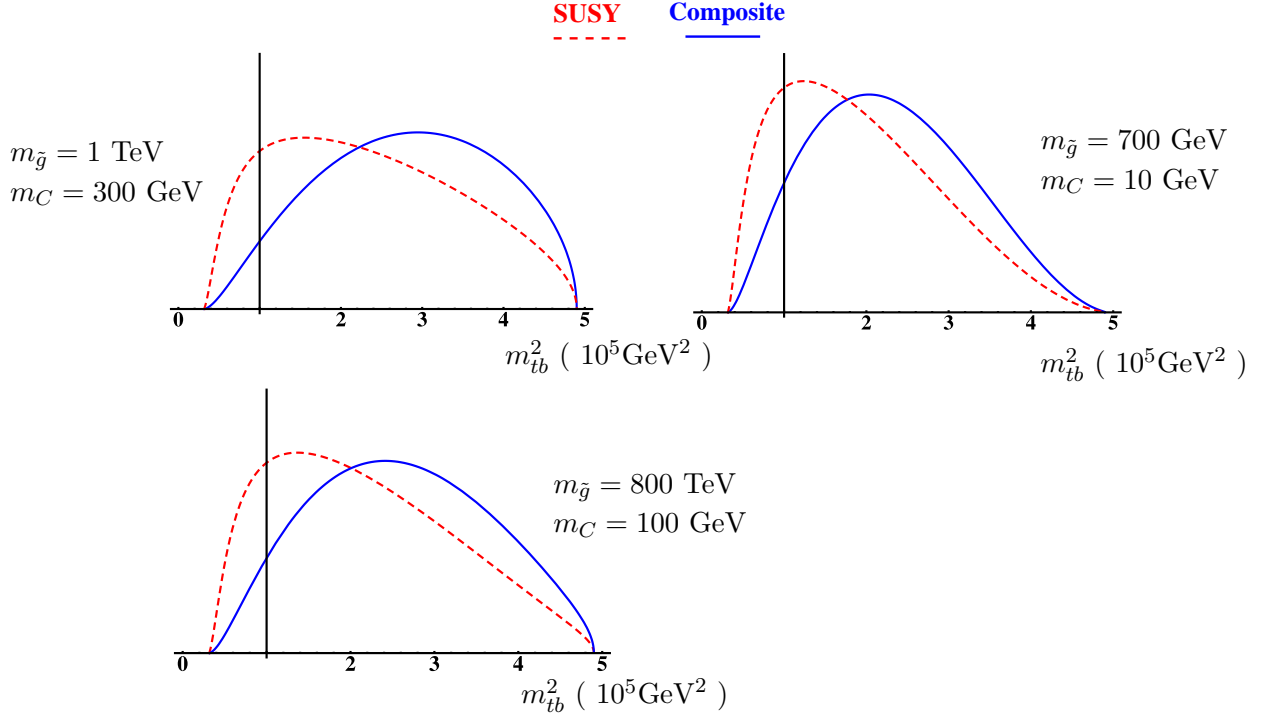}
\caption{$m_{tb}^2$ distribution for the operator $c_t$(in blue) and $c_{3}$(in red), for various spectrum all having the same end point.}
 \label{fig: mtbtheory2}
\end{center}
\end{figure}

The only invariant mass distribution relevant to the gluino decay that can be measured is $m_{tb}$. In figure \ref{fig: mtbtheory} and \ref{fig: mtbtheory2} we show the distribution of $m_{tb}^2$ given by the operators $c_t$ and $c_3$ for different spectra. We notice that the susy gluino ($c_t \neq 0$) distribution has a different shape than the composite gluino ($c_3 \neq 0$). The difference  comes from the fact that the expression for the differential width $d\Gamma/dm_{tb}^2$ starts with a constant 
in the supersymmetric gluino case and with a term proportional to $m_{tb}^2$ in the composite case.  The absence of a constant
in the composite case
is again a consequence of the charge conjugation symmetry. 
The differential widths are given by the following:
\begin{equation}
\frac{d\Gamma}{dm_{tb}^2} = c_3^2 \frac{1}{512 m_{\tilde{g}}^3  \pi^3 f^4} m_{tb}^2 (m_{C}^2+m_{\tilde{g}}^2 - m_{tb}^2) \sqrt{(m_{tb}^2-m_{C}^2)^2 - 2 (m_{tb}^2 + m_{C}^2) m_{\tilde{g}}^2 + m_{\tilde{g}}^4},
\end{equation}
for $c_t=c_b=c_4=0$ in the limit of zero $m_b$ and $m_t$ . While for $c_3=c_4=c_b=0$, we get
 \begin{multline}
 \frac{d\Gamma}{dm_{tb}^2} = c_t^2 \frac{1}{3072 m_{\tilde{g}}^3  \pi^3 f^4} \left( \left(m_{\tilde{g}}^2 - m_{C}^2 \right)^2 + m_{tb}^2 \left(m_{C}^2 + m_{\tilde{g}}^2 \right) -2 m_{tb}^4 \right) \\
 \sqrt{  \left(m_{\tilde{g}}^2 - m_{C}^2 \right)^2  - 2 m_{tb}^2 \left(m_{C}^2 + m_{\tilde{g}}^2 \right) + m_{tb}^4}
 \end{multline}
At small $m_{tb}^2$, these formulae change due to the non-zero value of $m_t$. In particular, both distributions  need to go to zero at $m_{tb}= m_t +m_b$, which is the kinematic limit.  The situation would be better if the final state was 2 $b$'s and a neutralino instead. The difference between the two distributions would be more pronounced  because the SUSY distribution would remain constant for lower values of $m_{bb}$. Unfortunately, there is no equivalent to operator $c_3$ for this final state, since the right-handed bottom is not expected to be composite. 

\section{Everything tastes like chicken: SUSY vs composite gluinos}
\label{results}
\begin{figure}[h]
\begin{center}
\begin{minipage}[b]{8.0cm}
\psfrag{LABEL}{\begin{tabular}{l} $m_{\tilde{g}}=1$ TeV \\ $m_C = 300$ GeV\\ $m_N = 200$ GeV \end{tabular}}
\includegraphics[width=8.0cm]{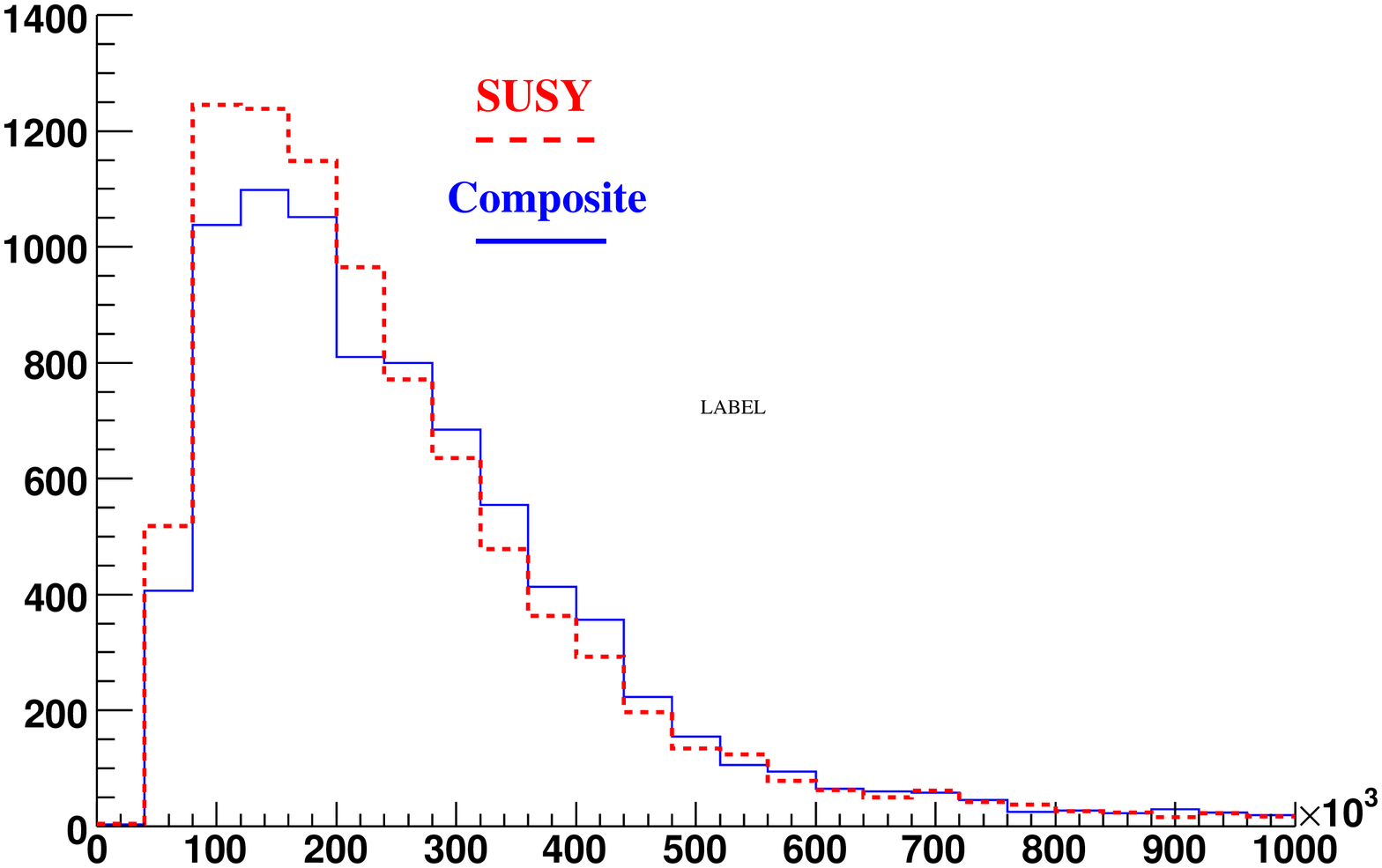}
\end{minipage}
\begin{minipage}[b]{8.0cm}
\psfrag{LABEL}{\begin{tabular}{l} $m_{\tilde{g}}=1$ TeV \\ $m_C = 100$ GeV\\ $m_N = 90$ GeV \end{tabular}}
\includegraphics[width=8.0cm]{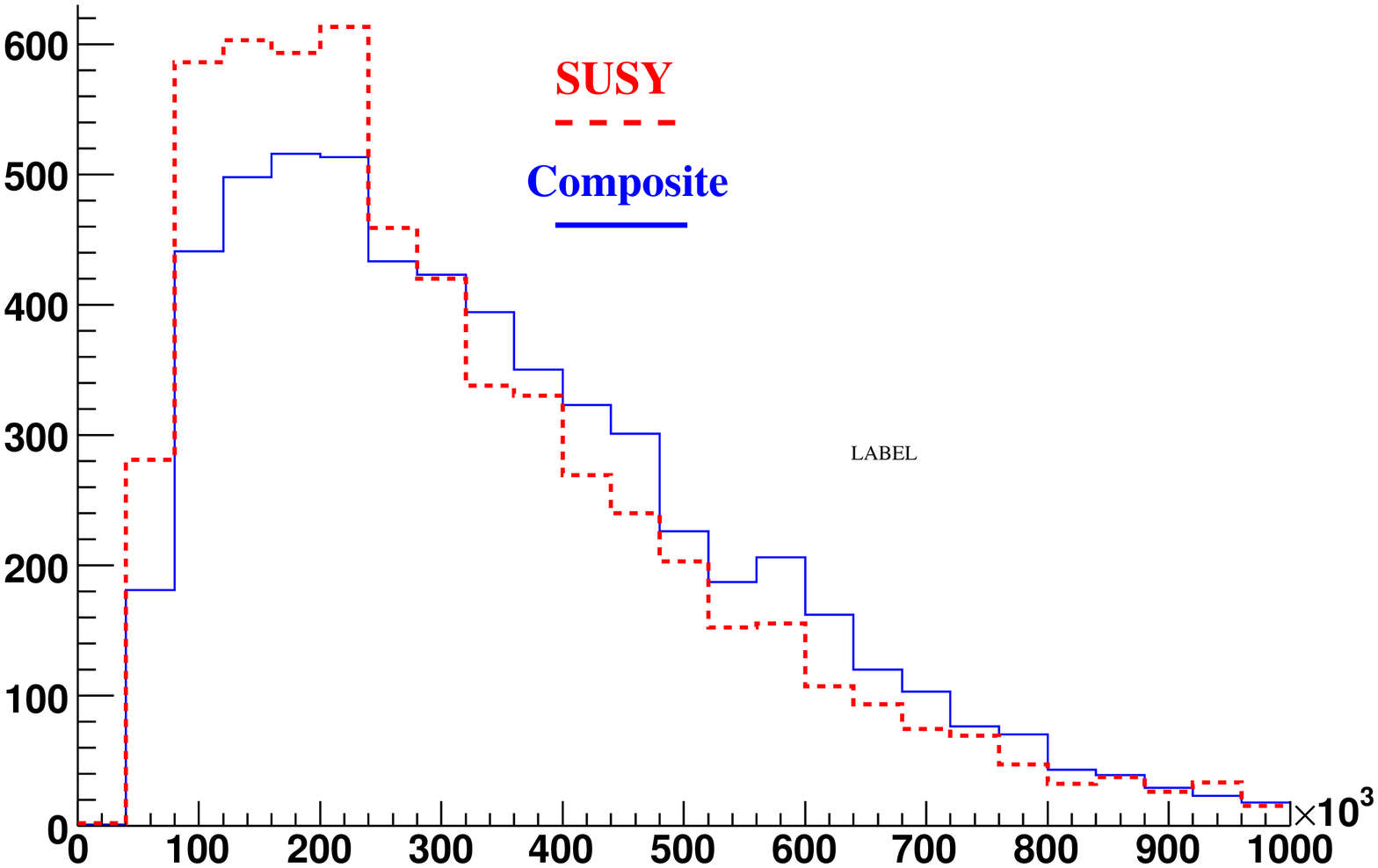}
\end{minipage}
\caption{ The blue and red histogram  are normalized to the same number of events. Composite corresponds to $c_4=1,c_3=c_t=c_b=0$, while SUSY corresponds to $c_t=1,c_b=c_3=c_4=0$.}
\label{fig: mtbsmallest}
\end{center}
\end{figure}

\begin{figure}[h]
\begin{center}
\begin{minipage}[b]{8.0cm}
\psfrag{LABEL}{\begin{tabular}{l}{\small $m_{\tilde{g}}=1$ TeV }\\{\small $m_C = 300$ GeV}\\{\small  $m_N = 200$ GeV }\end{tabular}}
\includegraphics[width=8.0cm]{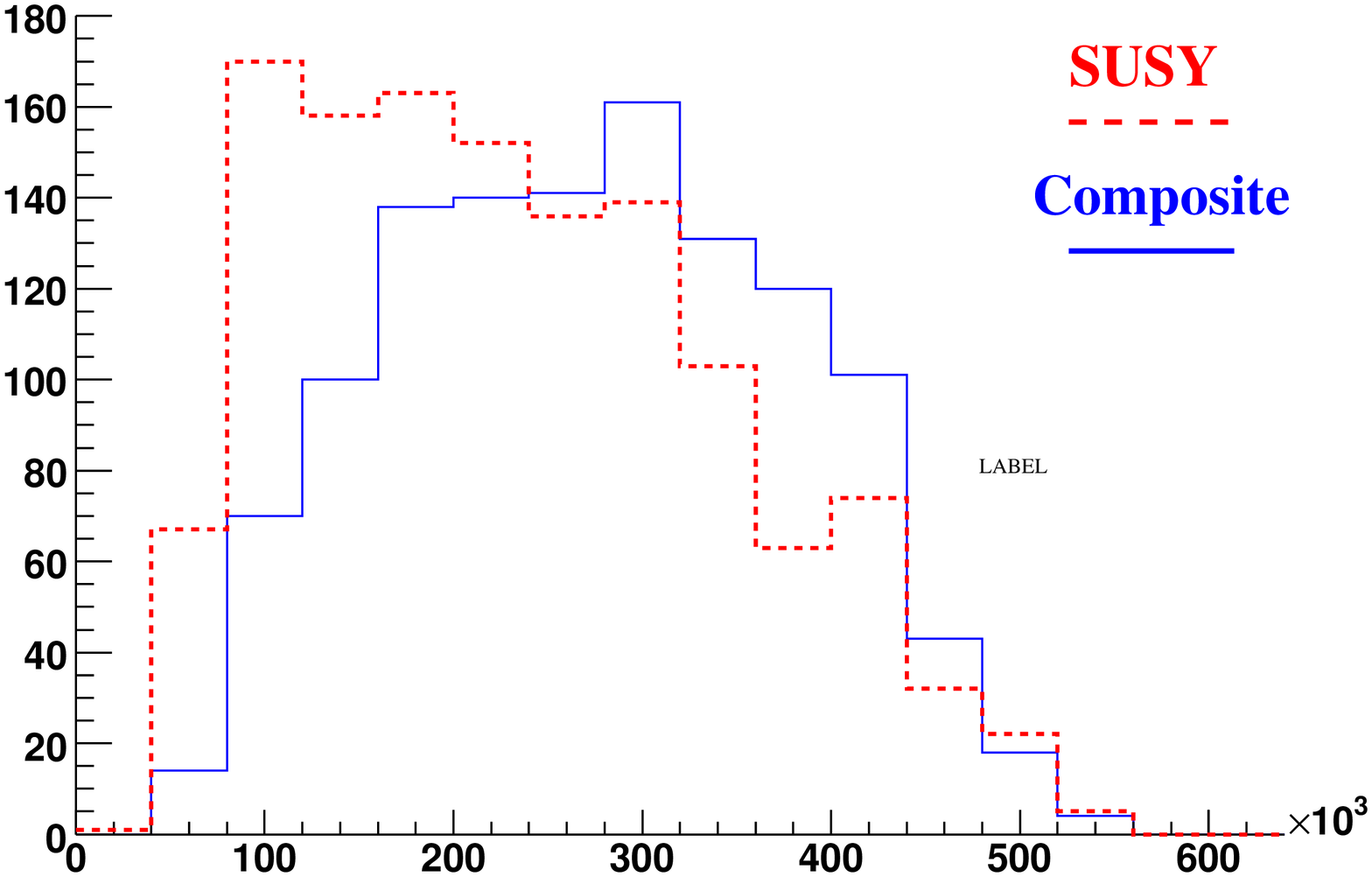}
\end{minipage}
\begin{minipage}[b]{8.0cm}
\psfrag{LABEL}{\begin{tabular}{l}{\small $m_{\tilde{g}}=1$ TeV }\\{\small $m_C = 100$ GeV}\\{\small  $m_N = 90$ GeV }\end{tabular}}
\includegraphics[width=8.0cm]{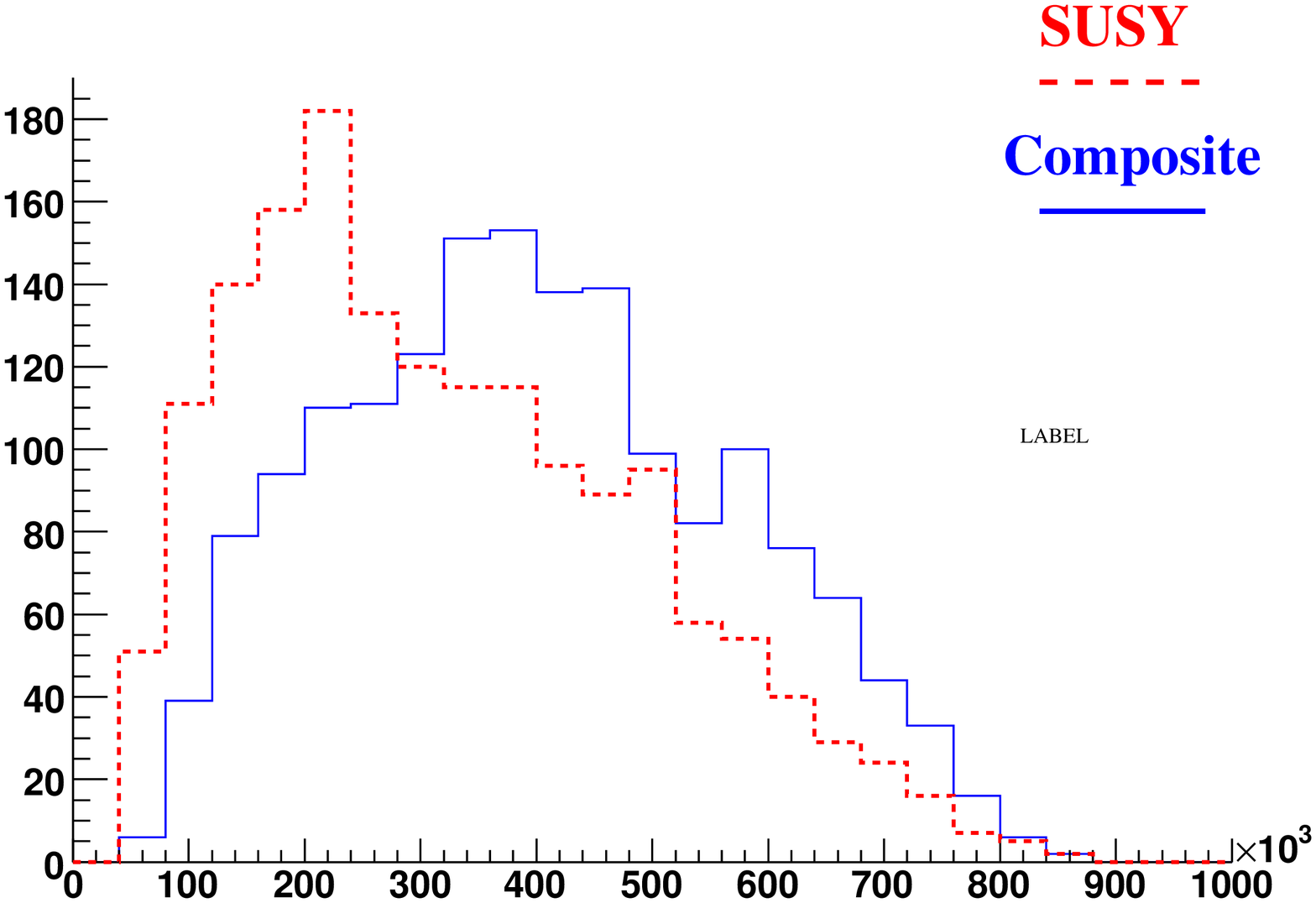}
\end{minipage}
\caption{$m_{tb}$ for events that were correctly reconstructed. The red and blue  histograms are normalized to the same number of events. Composite corresponds to $c_4=1,c_3=c_t=c_b=0$, while SUSY corresponds to $c_t=1,c_b=c_3=c_4=0$.}
\label{fig: mtbtrue}
\end{center}
\end{figure}

In principle, the two models could be distinguished by  precisely measuring the $m_{tb}$ distribution. However in practice things are obviously difficult. One needs to reconstruct the top through it's hadronic decay, and choose the right $b$ quark to pair it with. Moreover, the gluinos are pair produced resulting in a large multiplicity of particles, and the combinatorical background is 
very large. To estimate the size of this background we performed a parton-level analysis of this scenario.  Using Madgraph 4\cite{Alwall:2007st,Maltoni:2002qb}, we generated $100000$ gluino pairs, corresponding to a luminosity of about $300 \text{fb}^{-1}$ for a gluino mass of $1$ TeV. Each gluino was then decayed to a top, a bottom and a chargino. The chargino was subsequently decayed to a neutralino and a $W$.  We applied a smearing of 
$\sigma= 68\%/\sqrt{E}  \oplus 4.4\%$ to the final state quark energy to model detector resolution. We then applied to following cuts to our signal, designed to reduce Standard Model background:
\begin{itemize}
\item
$4$ or more jets(quarks) with $p_T > 40$ GeV and $\left|\eta \right| < 2.5$, two or more of which are b-tagged.
\item one or more jet(quark) with $p_T > 150$ GeV.
\item
missing $E_T$ greater than $100$ GeV.
\end{itemize}
where the various quarks have the following chance of being b-tagged: $50 \%$ for b-quarks, $10\%$ for charm quarks, 
and $1\%$ for light quarks (for $|\eta| < 1.5$). We note that the cut of $p_T > 40$GeV is probably too soft to avoid initial and final state radiation\cite{Plehn:2005cq}, and we expect the situation to be more complex in a realistic study.  However, a brief 
exploration of this issue indicates that a harder cut will not modify our results significantly.
We also did not apply smearing on the missing energy, as we are not including Standard Model backgrounds at this stage. 
To reconstruct our invariant mass distribution, we proceed as follows:
\begin{itemize}
\item We find two non-b-jets that reconstruct a $W$ : $ \left(m_W - 20 \text{GeV}\right)^2 < (p_1 + p_2)^2 < (m_W + 20 \text{GeV})^2$
\item We find a b-jet that reconstruct a top with the previously reconstructed $W$: $(m_t- 30 \text{GeV})^2 < (p_W + p_b)^2 < (m_t + 30 \text{GeV})^2$.
\item Out of all the reconstructed 'tops', we take as the 'right' top the one whose invariant mass is closest to the real mass of the top.
\item We find a 'hard'  b-jet with $p_T > 100$ GeV that we combine with the previously reconstructed top to reconstruct $m_{tb}$.
\item If there are different possible 'hard' b-jets, we choose the one that minimizes $m_{tb}$. 
\end{itemize}
We show the resulting $m_{tb}^2$ distribution in figure \ref{fig: mtbsmallest}. Even though the two histograms are statistically different, this will probably  not survive systematic uncertainty such as showering, hadronization and detector effects.  The ratio of correct reconstruction for this spectrum is around 15\% both for the composite and SUSY(with $c_t = 1$, $c_b=0$) gluino . Only about $60\%$ of the reconstructed $m_{tb}$ have a correctly reconstructed $W$'s, and $38\%$ have a correctly reconstructed top.  Moreover, the fraction of correctly reconstructed events over the number of generated events is about $1\%$. In figure \ref{fig: mtbtrue} we plot only the events where $m_{tb}$ was reconstructed correctly. The shapes are clearly distinguishable, showing that the cuts and smearing that we applied preserve the main features of the shapes, and the real issue is the combinatorical background. 

\begin{figure}[h]
\begin{center}
\psfrag{Spectrum2}{\begin{tabular}{l} $m_{\tilde{g}}=1$ TeV \\ $m_C = 100$ GeV\\ $m_N = 90$ GeV \end{tabular}}
\psfrag{Labelx}{$m_{tb}^2$}
\psfrag{Spectrum}{\begin{tabular}{l} $m_{\tilde{g}}=1$ TeV \\ $m_C = 300$ GeV\\ $m_N = 200$ GeV \end{tabular}}
\includegraphics[width=16.0cm]{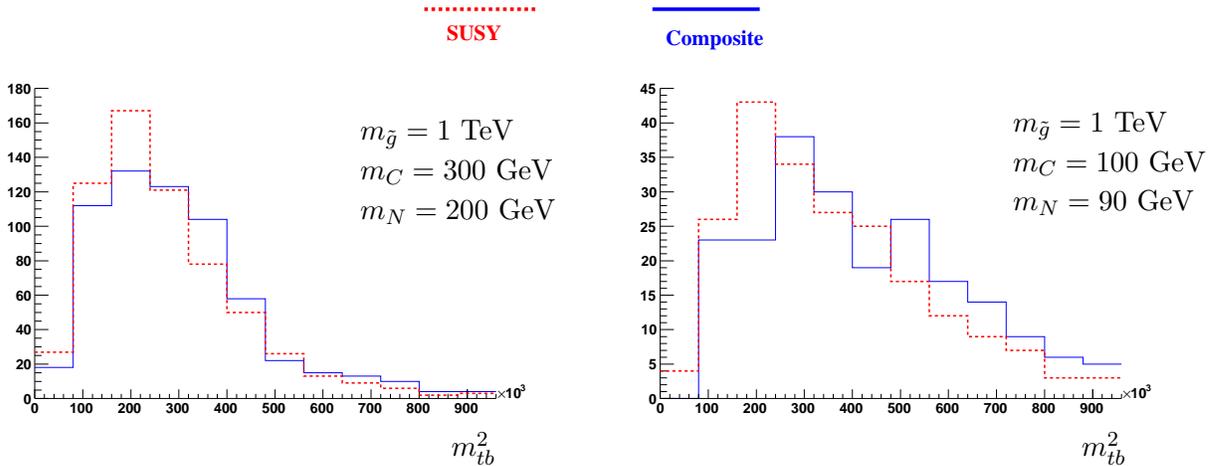}
\caption{ Results for two reconstructed tops. The composite label corresponds to $c_4=1,c_3=c_t=c_b=0$, while SUSY corresponds to $c_t=1,c_b=c_3=c_4=0$.}
\label{fig : 2top}
\end{center}
\end{figure}

By trying to reconstruct two tops instead of one, we can get a somewhat better purity at the expense of the number of events. We tried an analysis similar to the one used in \cite{AtlasTDRpoint5} to study 
MSUGRA point 5.  We required at least $7$ jets, 3 of which are b-tagged. We then looked for $2$ pairs of non-b jets which both reconstruct a $W$ (invariant mass within $20$ GeV of the $W$ mass). 
We then matched each pair with a b and kept the two $b-j-j$ combinations that minimized $\chi = ({m_1}_{bjj}-m_{\text{top}})^2 + ({m_2}_{bjj}-m_{\text{top}})^2$. Next, we required that the invariant mass of both combinations be within $30$ GeV of the top mass. These top candidates were then matched with the remaining $b$ jet which was required to have $p_T > 100$ GeV to construct $m_{tb}$. The percentage of correct reconstruction rises to around $20\%$ ($16\%$ in the composite case), but at the expense of the number of reconstructed events. The result is show in figure \ref{fig : 2top}.

Different spectra can lead to situations were the comparison is  slightly easier, for example, if the mass difference between the gluino and the chargino is larger, and the mass difference between the chargino and neutralino is smaller. The endpoint of the distribution, given by $(m_{\tilde{g}}-m_C)^2$, is then further away, and if  the chargino decays to the neutralino trough a three body decay, the extra (now off-shell) W is not reconstructed resulting in a reduced combinatorical background. The results for such a spectrum are shown on the right in figures \ref{fig: mtbsmallest} \ref{fig: mtbtrue}, and \ref{fig : 2top}. The fraction of correct reconstruction is $28\%$ for the analysis that reconstruct one top, and around $ 45 \%$ for the other analysis that reconstruct two tops. In the first analysis, the percentage of events with a correctly reconstructed top is now $73 \%$.  In figure \ref{fig: mtbtrue} we also plot  the correctly reconstructed events only in this analysis.
 
\begin{figure}[h]
\begin{center}
\begin{minipage}[t]{8cm}
\psfrag{LABEL1}{\begin{tabular}{l}{\footnotesize $m_{\tilde{g}}=800$ GeV}\\{\footnotesize $m_C = 100$ GeV}\\{\footnotesize  $m_N = 90$ GeV }\end{tabular}}
\psfrag{LABEL2}{\begin{tabular}{l}{\footnotesize $m_{\tilde{g}}=1$ TeV}\\{\footnotesize $m_C = 300$ GeV}\\{\footnotesize  $m_N = 200$ GeV }\end{tabular}}
\psfrag{LABELX}{$m_{tb}^2$}
\includegraphics[width=8cm]{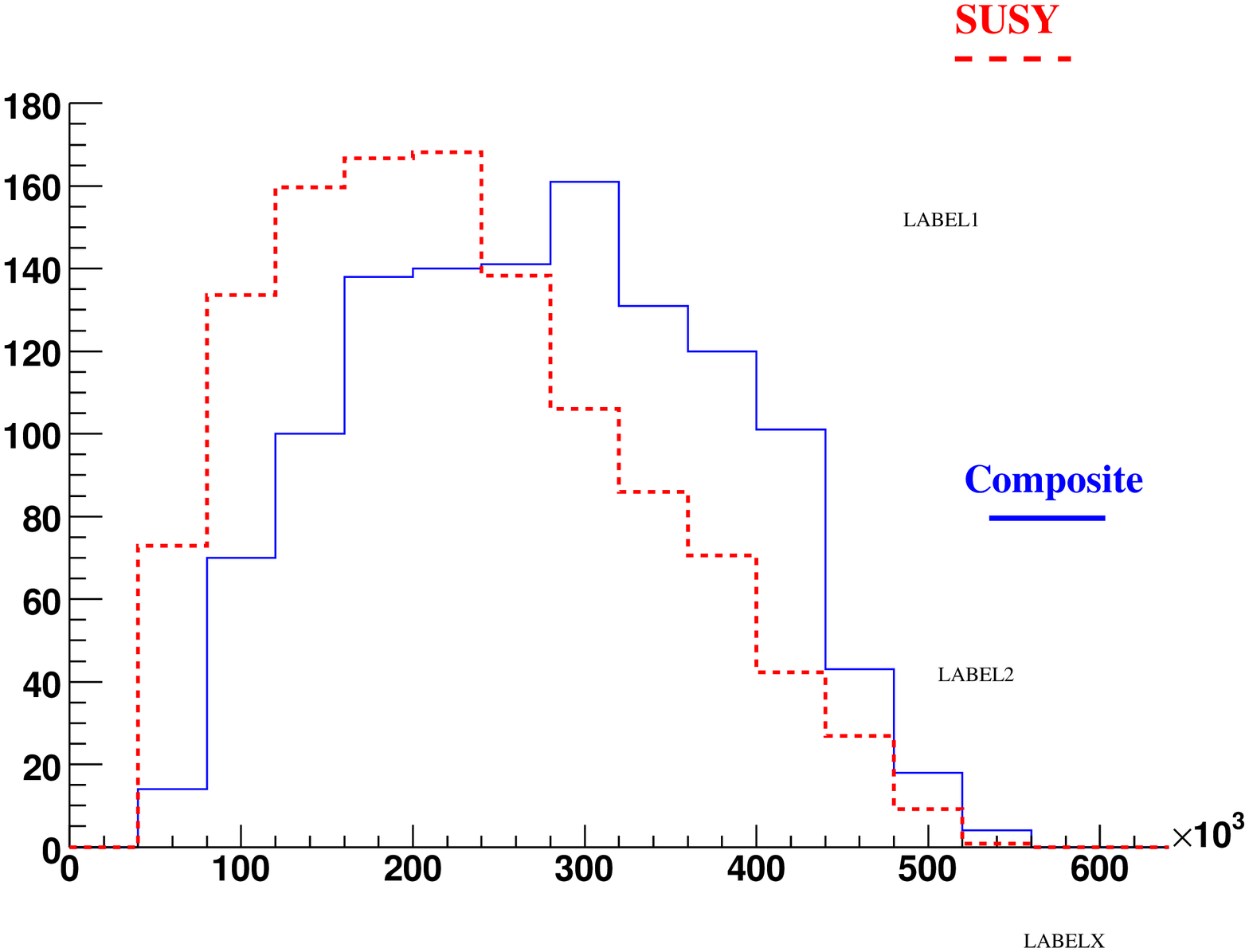}
\end{minipage}
\begin{minipage}[t]{8cm}
\psfrag{LABEL1}{\begin{tabular}{l}{\footnotesize $m_{\tilde{g}}=800$ GeV}\\{\footnotesize $m_C = 100$ GeV}\\{\footnotesize  $m_N = 90$ GeV }\end{tabular}}
\psfrag{LABEL2}{\begin{tabular}{l}{\footnotesize $m_{\tilde{g}}=1$ TeV}\\{\footnotesize $m_C = 300$ GeV}\\{\footnotesize  $m_N = 200$ GeV }\end{tabular}}
\psfrag{LABELX}{$m_{tb}^2$}
\includegraphics[width=8cm]{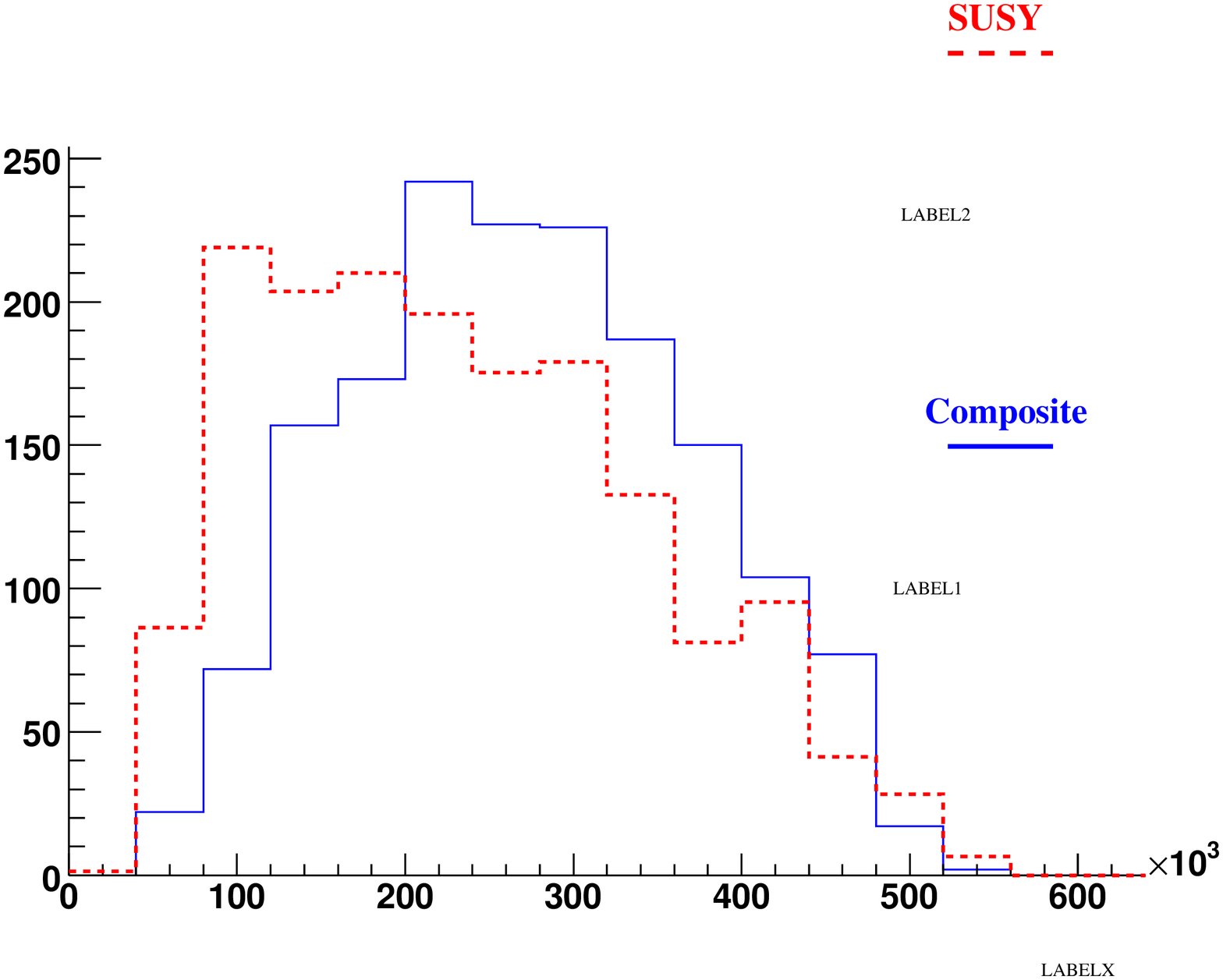}
\end{minipage}

\caption{$m_{tb}^2$ distribution(only the correctly reconstructed events), for composite and supersymmetric gluino with different spectrum. The two histograms are normalized to the same number of events. Composite corresponds to $c_4=1,c_3=c_t=c_b=0$, while SUSY corresponds to $c_t=1,c_b=c_3=c_4=0$. }
\label{fig: mtbdiffmass}
\end{center}
\end{figure}

The comparisons shown above were made for SUSY and composite 'gauginos' of the same mass. However, in practice,it might be hard to pin down the exact spectrum of the decay. In figure \ref{fig: mtbdiffmass} we look at the effect of comparing a composite and a supersymmetric gluino  with different spectra, but with the same end point(artificially removing the combinatorics).  We see that distinguishing  a composite gluino from a heavier supersymmetric gluino  is harder.  However, since the cross section for gluino pair production is very sensitive to the gluino mass, we expect to be able to determine the gluino mass to about $10\%$ \cite{Baer:2007ya}.

\subsection{Distribution with leptons}
\begin{figure}[h]
\begin{center}
\psfrag{Spectrum}{\begin{tabular}{l} $m_{\tilde{g}}=1$ TeV \\ $m_C = 300$ GeV\\ $m_N = 200$ GeV \end{tabular}}
\psfrag{Labelx}{$m_{lb\bar{b}}$}
\includegraphics[width=15cm]{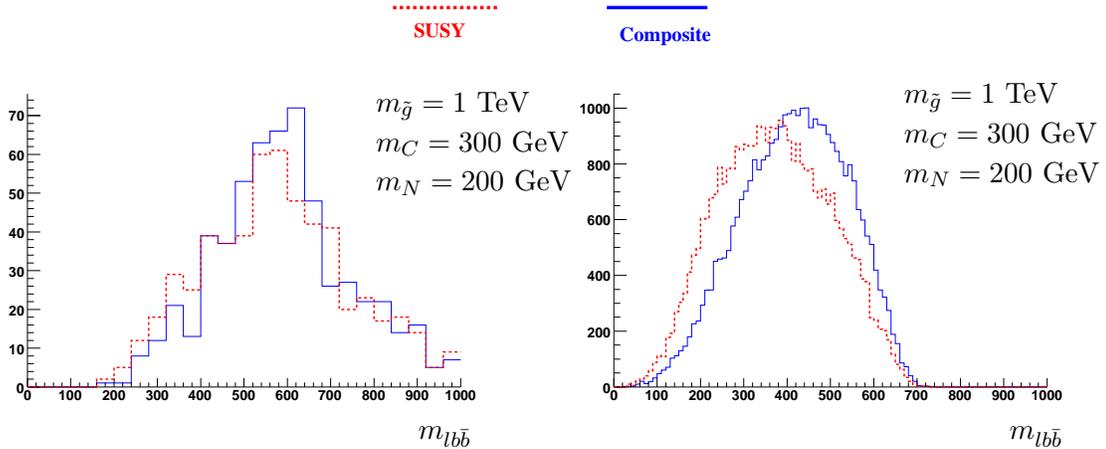}
\caption{$m_{bbl}$ distribution. The plot on the right shows the exact distribution for the composite and SUSY case (with quark energy smearing). On the left the plot shows the same comparison with combinatorical background. }
\label{fig : lepton}
\end{center}
\end{figure}

In the previous subsection, we found that the $m_{tb}$ distribution could in principle be used to discriminate between a supersymmetric theory and a composite one. However, reconstruction of this observable is extremely challenging. Even without including extra jets from initial and final state radiation, underlying events, jet reconstruction, etc., the combinatorics are such that we do not effectively discriminate between 
our two competing scenarios. It is therefore interesting to consider observables that include leptons, which are experimentally cleaner. These observables tend however to be more similar in the 
two scenarios. We found that the most promising candidate as a discriminant is $m_{b \bar{b} l} = \left|p_b + p_{\bar{b}} + p_l \right|$, where the lepton, $l$, and one of the $b$'s comes from the top, while the other $b$
comes the gluino decay.  This distribution is shown in figure \ref{fig : lepton}.
Unfortunately, reconstruction of this observable is also subject to a large combinatorical background. Without additional kinematic discriminants, only a third of the possible $b \bar{b}$ pairs are the correct
ones.  Moreover, once we pick one such pair, there are $4$ possible leptons. 

In an attempt to reduce the combinatorical background we identified  the $b \bar{b}$ pairs as the highest and lowest $p_T$ b-tagged jets in each event. Furthermore we required that the first jet had $p_T>250$ GeV, and the second  $p_T < 100$ GeV. We also  required  the invariant mass of the $b$ pair to be less than $600$ GeV. To improve identification of the lepton, we required that it had $p_T>150$ GeV, and we picked the one that  closest to the lowest $p_T$ b-tagged jet in the $\phi-\eta$ plane. We finally required that it's distance to the lowest $p_T$ b-tagged jet  be less than 3 in the $\phi-\eta$ plane and  it's invariant mass with this same jet be less than $175$ GeV. This achieves a $\sim 40 \%$ purity for the $b\bar{b}$ pairing, and $\sim 20\%$ for the final $b-\bar{b}-l$ combination. Results are showed in figure \ref{fig : lepton}. We also show the exact distribution( with quark energy smearing).

\subsection{Backgrounds}
Given the very small differences in shape found in the previous subsections, we do not attempt the same comparison in a more realistic environment including parton showering, hadronization and the presence of 
Standard Model backgrounds. However, we can ask if our composite and SUSY gluinos can be seen clearly above the background. There have been various studies of signals closely resembling ours in the 
context of focus point supersymmetry \cite{Baer:2007ya,DeSanctis:2007td,Das:2007jn,Mercadante:2005vx,Baer:2005ky}. There, the gluinos are quite heavy, and the squarks are even heavier, with the stop lighter than the rest. Thus, the glunio decays through an off-shell stop 
to $t \bar{b} C^-$ or $t \bar{t} N$, just as in our scenario.

The various studies \cite{Baer:2007ya,DeSanctis:2007td,Das:2007jn,Mercadante:2005vx,Baer:2005ky} for focus point supersymmetry found that with about $10 \text{fb}^{-1}$ of data,  the signal can be easily distinguished from Standard Model background.
The background consists mostly of top and bottom pair production which was reduced by requiring a large number of jets, large missing energy, many b-tagged jets, and a large effective mass. 
We verified that it was also the case for our particular signal by showering and hadronizing the parton-level events with Pythia\cite{Sjostrand:2006za} and using PGS\cite{PGS} to model detector effects. We also generated a
$t \bar{t}$ and $t \bar{t} +1$  jet  sample using Alpgen\cite{Mangano:2002ea} for the parton level production, Pythia\cite{Sjostrand:2006za} for showering and hadronizing and PGS\cite{PGS} as a detector simulation.  By using  cuts very similar to \cite{Baer:2007ya}, namely asking for more than $7$ jets 
with $P_T > 40$ GeV,  $\sla{E}_T > 100$ GeV, 2 b-tagged jets and $A_T = \sla{E}_T + \sum_{\text{jets,leptons}}p_T > 1400$ GeV, we find that the gluino signal can be isolated from the background. With $100 \text{fb}^{-1}$ we get $S/\sqrt{B}$ of 
over 68 and $S/B \sim 0.8$ (considering only the $t\bar{t}$ background). The situation is even better if we require a reconstructed  $m_{tb}$ invariant.

\section{On-shell stop}
\label{stop}
The supersymmetric model considered in the previous section was specifically designed to fake our composite gluino. But, since it may be justified to have the third generation special, and lighter than the rest of the squarks and sleptons, the superpartner of the top, the stop, might not be very heavy. Since the stop has no equivalent in the composite model, directly observing it would  point towards a supersymmetric signal. This would be impossible if the stop is heavier than the gluino, and too heavy to be significantly pair produced.  However, if it's light enough, it could be produced on-shell, either directly or through gluino decay. 

\begin{figure}[h]
\begin{center}
\psfrag{Labelx}{$m_{bW}$}
\psfrag{Spectrum}{ $m_{\tilde{g}}=1$ TeV, $m_C = 300$ GeV, $m_N=200$ GeV}
\psfrag{Spectrum2}{($m_{\tilde{t}} = 500$ GeV)}
\includegraphics[width=15cm]{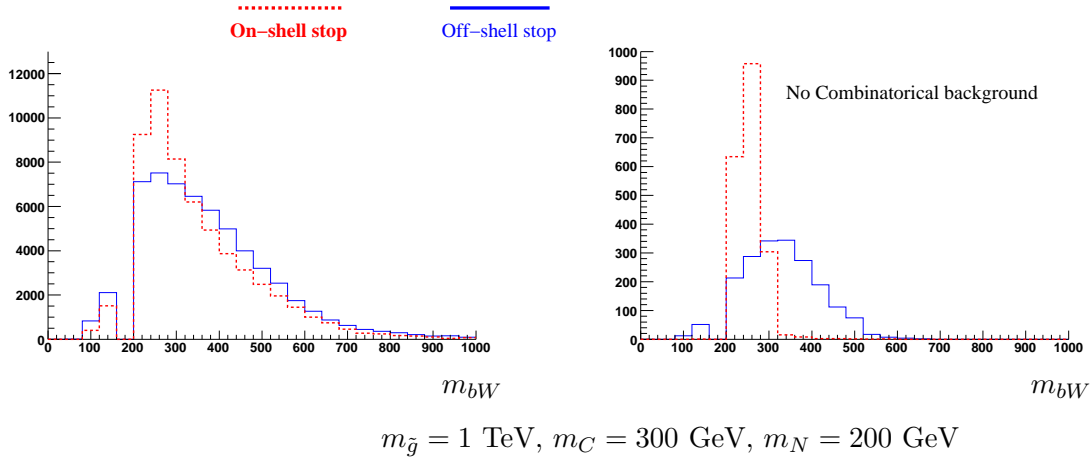}
\caption{$m_{bW}$ distribution for events with on-shell stops, compared to events with off-shell stop . The mass of the stop is $500$ GeV in the on-shell case.}
 \label{fig: wb}
\end{center}
\end{figure}

If we assume that the stop always decays through $\tilde{t} \rightarrow b C^+ \rightarrow b W^+ N$, the presence of an on-shell stop in the decay chain would  introduces more structure in the $m_{bW}$ invariant mass distribution. It would then have an edge at (neglecting the $W$ and $b$ mass):
\begin{equation}
{m^2_{bW}}_{\text{edge}} = \frac{\left(m^2_{\tilde{t}} -m^2_{C}\right) \left(m^2_C - m^2_N \right)}{2 m_C^2} \; .
\end{equation}
Seeing this endpoint would be an indication for the presence of a stop. In figure \ref{fig: wb} we show the $m_{bW}$ invariant mass distribution for events containing on-shell stop (coming from gluino decay or stop pair production) compared with the same distribution when the stop is off-shell. The $W-b$ pairs are chosen as in the previous section, but outside the top mass rage: $|m_{bW}-m_{\text{top}}| > 30$ GeV. We see that this distribution is also plagued by large combinatorial backgrounds.

Another possibility is to try to isolate stop pair production. This has the same finale state as top pair production which ends up being  the main Standard Model background once a b-tagging cut is applied (see \cite{CMSnote} for example).  To reduce this background, we applied cuts similar to the one used in \cite{CMSnote}.  We first  asked for $\sla{E}_T > 100$  GeV, 4 or more jets with $p_T>40$ GeV, one of which is b-tagged and one and only one lepton. This selects one hadronic and one leptonic decay of the $W$'s in the events. We then put a cut on the transverse mass: ${m_{l \nu}}_T = \sqrt{2 {E_l}_T \sla{E}_T(1-\cos \phi{l \nu})} > 110$ GeV\cite{CMSnote}. which removes a great fraction of $t\bar{t}$ events for which ${m_{e \nu}}_T$ has a maximum near the $W$ mass. In the signal events, there is missing energy from the neutralino and the transverse mass can be much larger. We find that with $100 \text{fb}^{-1}$ of data, a $500$ GeV stop, and a $1$ TeV gluino, we can obtain $S/\sqrt{B} \sim 6.6$, where the background contains the gluino events and the $t\bar{t}$($+ 1$ jet) sample generated as in the previous section. 
It is also possible to identify the stop if it decays to a top and a neutralino \cite{ATLASnote,CMSnote}. In that case, one can reconstruct two tops in an event with large missing $E_T$. Again, SUSY background from the gluino decay are large, but can be dealt with by applying cuts on the hardness of the jets.

\section{Conclusions}
\label{end}
We have explored the possibility of distinguishing a supersymmetric gluino from a composite fermion with the same quantum numbers coming from a model where the Higgs and the top are composite. The composite gluino will decay through four-fermi interactions. Because of the composite nature of the third generation in these models, and because of an $R$-parity, under which the gluino is odd, the gluino will decay to two third generation fermions and one stable composite 'neutralino'. We chose to study only the decay $\tilde{g} \rightarrow t \bar{b} {C^-} \rightarrow t \bar{b}  {W^-}^{(*)} N$, with a 'higgsino'-like chargino and ignored a possible complication from decay to two tops and a neutralino which might very well be relevant in specific models.  These final states could also easily arise in a supersymmetric model where the sleptons and the first two generations of squarks are heavy. We found that the four-fermi interactions of the composite model have a form that could be distinguished from most supersymmetric models by examining the shape of the distribution of the top-bottom invariant mass, $m_{tb}$. However, this distribution, that could be measured 
through the hadronic decay of the top, is plagued by large combinatorical backgrounds. Cleaner observables that include leptons are more similar in the composite and supersymmetric models. The situation improves if the supersymmetric stop is light enough. Since this particle does not apriori exist in the composite model, observing it would be powerful evidence for supersymmetry. The stop could be observed  directly from stop pair production. There are large backgrounds from top pair production and from gluino decay, but we found that they could be overcome for a stop of $m_{\tilde{t}} \sim 500$ GeV and and gluino of  $m_{\tilde{g}} \sim 1$ TeV.  This study shows that observing a gluino might not be a direct indication of supersymmetry, and new techniques might be required to probe the structure of its various decays.  
\section*{Acknowledgment}
We thank Tilman Plehn, Michael Peskin and Veronica Sanz for helpful discussions. We also thank T. Plehn for comments on the draft.
E.K. was supported in part by the Department of Energy grant no. DE-FG02-01ER-40676, by the NSF CAREER grant PHY-0645456, 
and by the Alfred P. Sloan Fellowship. TG is supported by a SUPA advanced fellowship.

\end{document}